\documentstyle[floats,twocolumn,aps,psfig,prl]{revtex}

\begin{document}
\draft

\twocolumn[\hsize\textwidth\columnwidth\hsize\csname @twocolumnfalse\endcsname

\title{CsMn(Br$_x$I$_{1-x}$)$_3$: Crossover from an $XY$ to an Ising
Chiral Antiferromagnet}

\author{R. B\"ugel$^1$, J. Wosnitza$^1$, H. v. L\"ohneysen$^1$,
T. Ono$^2$, H. Tanaka$^2$}

\address{$^1$Physikalisches Institut, Universit\"at Karlsruhe,
D-76128 Karlsruhe, Germany\\
$^2$Department of Physics, Tokyo Institute of Technology, Tokyo 152,
Japan
}

\date{\today}
\maketitle

\begin{abstract}
We report on high-resolution specific-heat and magnetocaloric-effect
measurements of the triangular-lattice antiferromagnets
CsMn(Br$_x$I$_{1-x}$)$_3$ with different $x$. The evolution of the
magnetic phase diagrams from the easy-axis system for $x = 0$ to the
easy-plane system for $x = 1$ was studied in detail. The specific-heat
critical exponent $\alpha$ of the almost isotropic $x = 0.19$ system
agrees with the value predicted for a chiral Heisenberg scenario.
In an applied magnetic field ($B = 6$~T) a crossover to a weak
first-order transition is detected.
\end{abstract}
\vskip2pc]

\section{Introduction}
The triangular-lattice antiferromagnets ABX$_3$ with CsNiCl$_3$ structure,
where the magnetic B$^{2+}$ ions form a triangular lattice, exhibit
frustration due to the antiferromagnetic interactions on a triangular
plaquette if the magnetic moments have a component in the triangular
$ab$ plane. The magnetic moments then form a 120$^{\circ}$ structure,
with the extra two-fold degeneracy of chirality being broken at the
antiferromagnetic transition. Simply speaking, the extra degeneracy
arises from the possibility that the 120$^{\circ}$ spin structure on
a given plaquette can be arranged clockwise or counterclockwise when
moving around the plaquette. It has been suggested that the chiral
degeneracy leads to new universality classes for three-dimensional $XY$
and Heisenberg models \cite{kaw85,kaw98}. The largest changes of the
critical exponents are predicted to occur in the specific-heat exponent
$\alpha$, where the chiral $XY$ and chiral Heisenberg universality
classes are predicted to show $\alpha = 0.34$ and 0.24, compared to
$\alpha = -0.01$ and $-0.12$ for the standard $XY$ and Heisenberg
models, respectively \cite{kaw98}. However, whether this concept of
new universality classes is indeed applicable is a strongly debated
question. Especially within recent years theoretical studies have
supplied growing support for a weakly first-order scenario for both
chiral phase transitions (see \cite{loi98,tis00,loi00} and references
therein). An experimental indication for this behavior was found
recently \cite{web96}.

In any case, the frustration enhances the degeneracy giving rise to
different physics with rich phase diagrams and strongly modified critical
behavior, which has been studied experimentally for a large number of
different triangular-lattice antiferromagnets \cite{col97}.
A well-studied example is the easy-plane system CsMnBr$_3$, for which
a number of experiments \cite{kad88,mas89,wan91,deu92,pla00} revealed a
critical behavior in line with the theoretical prediction \cite{kaw98}.
For ABX$_3$ systems with easy-axis anisotropy like CsMnI$_3$ (as well
as CsNiCl$_3$), chiral behavior can be induced by applying a
spin-flop field along the easy $c$ direction thus forcing the
spins into the $ab$ planes. At the spin-flop field $B_M$
($\sim$6.4~T for CsMnI$_3$ and $\sim$2.3~T for CsNiCl$_3$) the
magnetic energy is equal to the anisotropy energy, i.e., full isotropy
in spin space is attained and chiral Heisenberg behavior is found
\cite{web95,loe95,bec93b,end94}. For higher fields, an easy-plane
anisotropy is induced and $XY$ chirality occurs \cite{bec93b,end94}.

Not many triangular-lattice antiferromagnets with negligible
anisotropy exist. Besides the above-mentioned materials at their
spin-flop fields, only the hexagonal antiferromagnet VBr$_2$ is
known. Indeed, for VBr$_2$ critical exponents were found in line
with the behavior predicted for a chiral Heisenberg system
\cite{wos94}. The possibility to tune a chiral Heisenberg
system is offered by the solid solution CsMn(Br$_x$I$_{1-x}$)$_3$
which spans the range from an easy-axis system ($x = 0$) to an
easy-plane system ($x = 1$). This system, therefore, allows
to study the crossover in the magnetic phase diagrams and its
influence on the critical behavior. In particular, the composition
with $x = 0.19$ presents an almost isotropic system and should
therefore follow chiral Heisenberg behavior \cite{ono98}.
Magnetization measurements of CsMn(Br$_x$I$_{1-x}$)$_3$ which gave
some information on the magnetic phase diagrams have already
been reported by Ono et al.\ \cite{ono98}. Here we report on detailed
specific-heat and magnetocaloric-effect measurements.

The spin-Hamiltonian that describes the system is given by
\begin{eqnarray}
{\cal H} & = & -J_c \sum_{i,j}^{\rm chain} {\bf S}_i \cdot {\bf S}_j -
J_{ab} \sum_{i,j}^{\rm plane} {\bf S}_i \cdot {\bf S}_j \nonumber \\
&& {} + D \sum_i(S_i^z)^2 - g\mu_B \sum_i {\bf B} \cdot {\bf S}_i.
\end{eqnarray}
The summation ($i,j$) is over nearest neighbors, with the first sum along
the $c$ direction and the second sum in the $ab$ plane, with the exchange
constants $J_c$ and $J_{ab}$, respectively. $D< 0$ corresponds to an
easy-axis system, $D>0$ to an easy-plane system. For CsMnBr$_3$, $J_c =
-0.89$~meV, $J_{ab} = -1.7~\mu$eV, and $D = 12~\mu$eV, for CsMnI$_3$,
$J_c = -1.5$~meV, $J_{ab} = -7.6~\mu$eV, and $D = -3.8~\mu$eV.
For both materials we are dealing with the $S = 5/2$ spins of Mn$^{2+}$.

\begin{figure}[ht]
  \centerline{\psfig{file=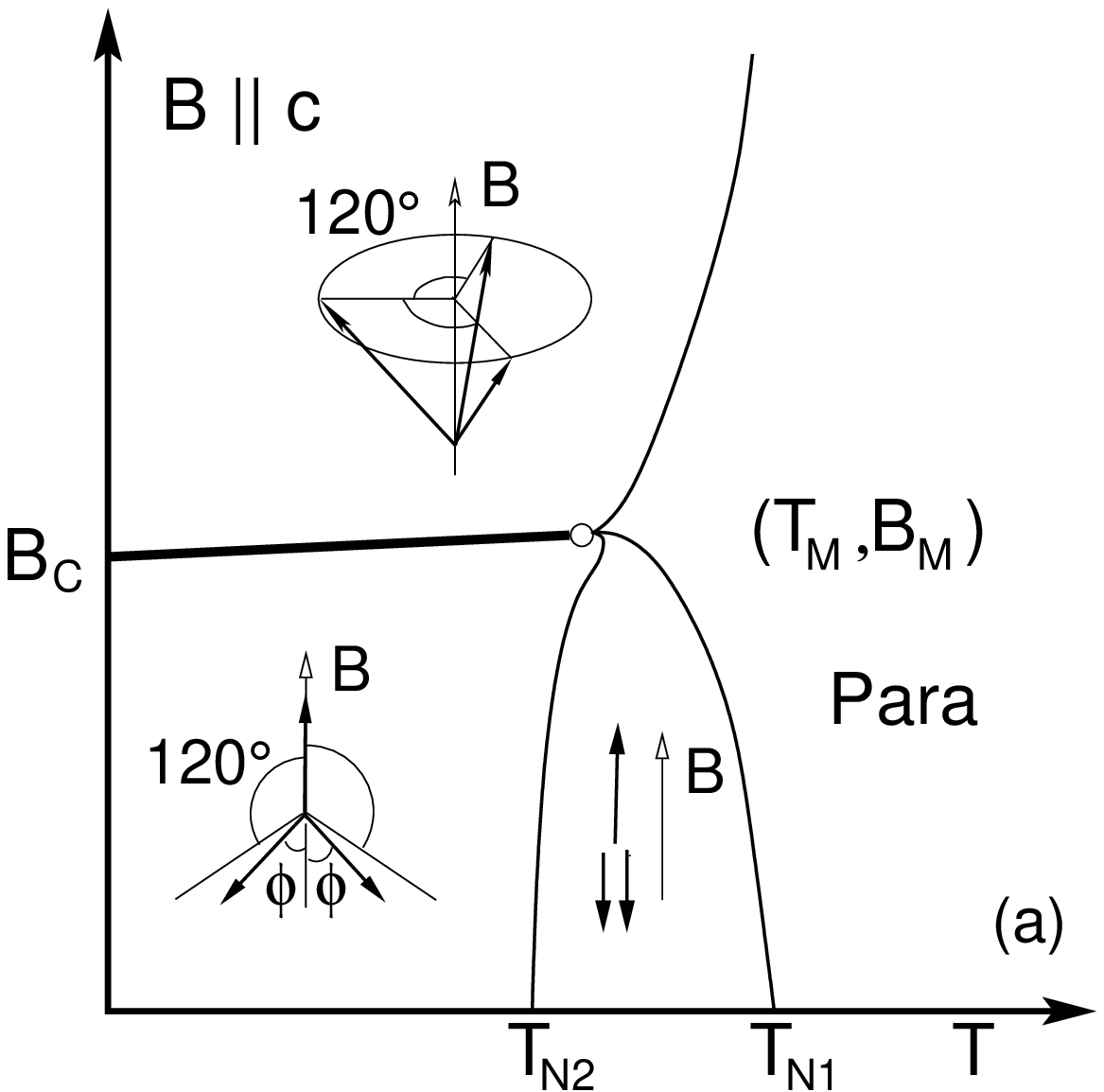,clip=,width=4.25cm}
  \psfig{file=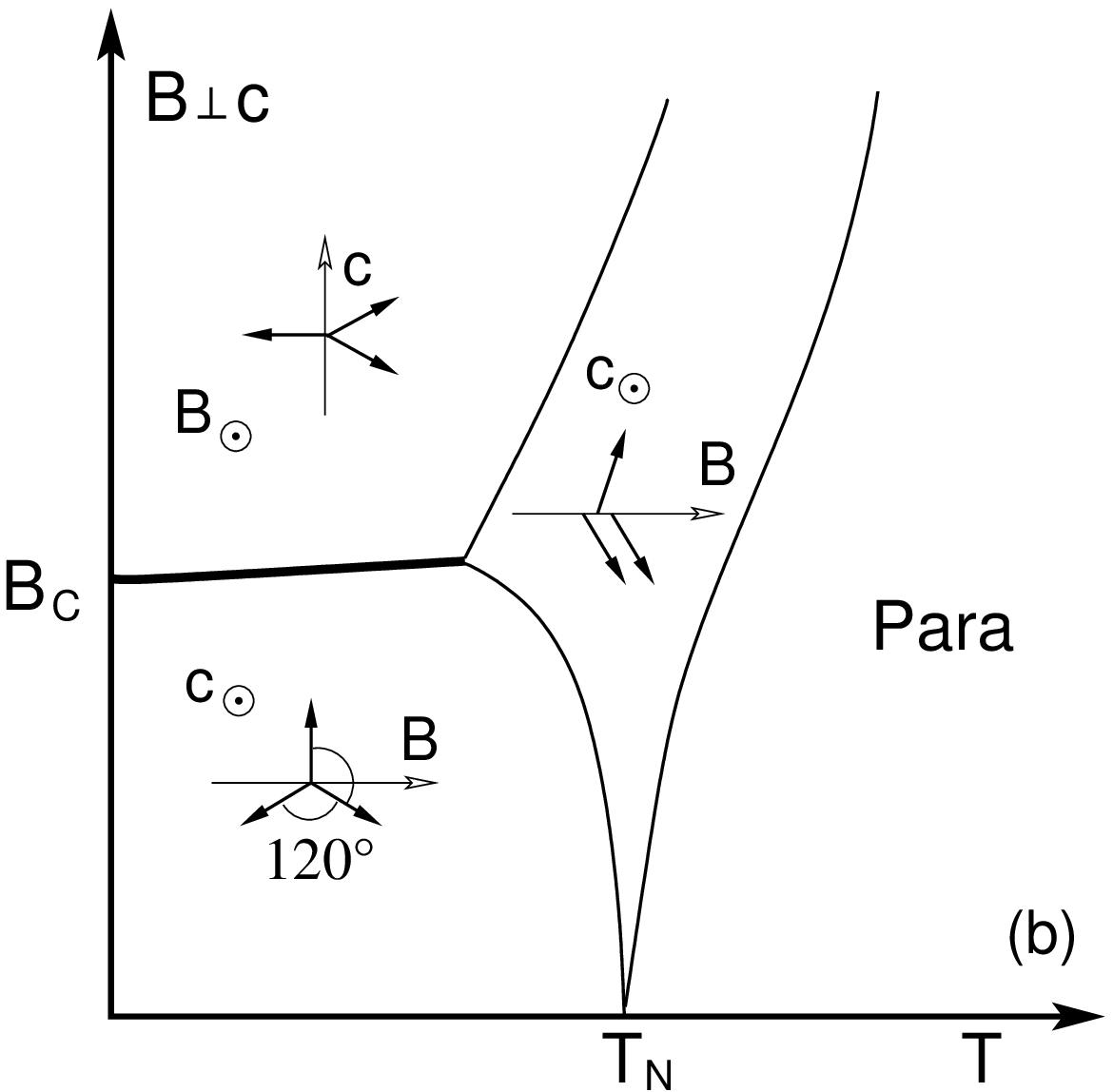,clip=,width=4.25cm}}
\caption[]{(a) Schematic phase diagram of a triangular-lattice
antiferromagnet with easy-axis anisotropy ($D < 0$) for $B || c$.
The phase lines meet at the multicritical point ($T_M,B_M$). $B_c$
is the critical field for the spin-flop transition. (b) The schematic
phase diagram for systems with a small easy-plane anisotropy $0 < D
< 3|J_{ab}|$ for $B \perp c$. The insets sketch the spin arrangements.}
\label{pds}
\end{figure}

The topology of the magnetic phase-diagrams for ABX$_3$ antiferromagnets
depends crucially on the sign of $D$ and on the ratio $D/J_{ab}$.
Systems with Ising anisotropy ($D< 0$) show two successive phase
transitions at $T_{N1}$ and $T_{N2}$ for $B = 0$. In the
low-temperature phase ($T < T_{N2}$) the spins order in three
sublattices where one third of the spins align along the $c$ axis,
whereas the other two thirds are tilted by an angle $\Phi$
(which depends on the ratio $D/J_{ab}$ \cite{col97})
with respect to the $c$ axis [see Fig.\ \ref{pds}(a)]. For CsMnI$_3$
this angle is $\Phi = 51^\circ$ \cite{har91}. All spins lie within
a plane which includes the $c$ axis. At higher temperatures in the
intermediate phase ($T_{N2} < T < T_{N1}$) the tilted spins have an
additional degree of freedom, i.e., their components within the
basal $ab$ plane is not defined. For $T > T_{N1}$ in the paramagnetic
phase only short-range ordered independent spin chains exist along
the $c$ axis. For a magnetic field applied within the basal plane
($B \perp c$) the two phase boundaries shift somewhat to higher
temperatures (at least up to 6~T for CsMnI$_3$ and CsNiCl$_3$)
without changing the principal spin topology \cite{web95,bec93a}.

Of much more relevance is the case when $B$ is applied along $c$
[Fig.\ \ref{pds}(a)]. For $T < T_{N2}$, a first-order phase
transition occurs at the spin-flop field $B_c$ above which the three
sublattices form a $120^\circ$ umbrella-like structure. The $c$-axis
spin component grows with further increasing $B$. For classical
spins with $J_c \gg J_{ab}$, $B_c$ at $T = 0$ is given by
\begin{equation}
(g\mu_BB_c)^2 = 16|J_cD|S^2.
\label{spinflop}
\end{equation}
At the multicritical point ($T_M$,$B_M$) the three phase lines merge
tangentially into the first-order spin-flop line \cite{plu88,kaw90}.
At this point full isotropy in spin space is achieved which leads to
a chiral Heisenberg universality as experimentally observed for
CsNiCl$_3$ \cite{bec93b,end94} and CsMnI$_3$ \cite{web95}.

For systems with easy-plane anisotropy ($D > 0$) only one phase
transition at $T_N$ from the paramagnetic to the chiral 120$^\circ$
structure exists at $B = 0$. The critical behavior, therefore, is
of the chiral $XY$ type. A magnetic field applied along the
$c$ direction does not change the symmetry of the ground state.
Consequently, the critical behavior stays essentially constant
\cite{deu92}. A much richer phase diagram can be observed for
$B \perp c$. The phase diagram as predicted for $D < 3|J_{ab}|$
is shown in Fig. \ref{pds}(b) \cite{plu89}. At $T < T_N$ and
$B < B_c$ the chiral phase exists. The competition between the
Zeeman energy and the anisotropy energy leads to a spin-flop phase
above $B_c$ which is also given by Eq.\ (\ref{spinflop}) \cite{col97}.
Thereby, the spin triangle is oriented perpendicularly to $B$. Between
the chiral low-temperature and the paramagnetic high-temperature phase,
a collinear spin structure evolves where the spins remain in the
$ab$ plane with two spins of a triangle pointing parallel and one
in the opposite direction. With increasing field the spin components
along $B$ become larger.

Experimentally, very little is known about the phase diagrams and the
critical properties of easy-plane systems with such a small anisotropy
($D < 3|J_{ab}|$) \cite{col97}. On the other hand, for materials with
$D > 3|J_{ab}|$, like CsMnBr$_3$, the phase diagram has been very
well established \cite{mas89,web95,gau89}. Due to the stronger
anisotropy $D$ the spin-flop phase is absent for such $XY$ systems
with only the chiral phase and the collinear structure remaining
\cite{col97}.

\begin{figure}[ht]
  \centerline{\psfig{file=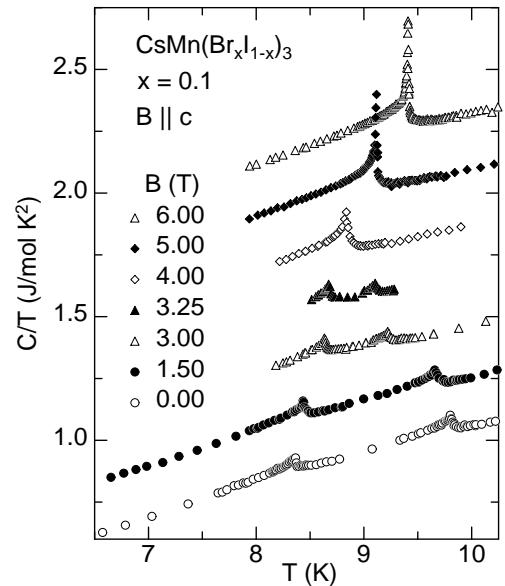,clip=,width=7cm}}
\caption[]{Specific heat $C$ divided by temperature $T$ vs $T$ for
CsMn(Br$_{0.1}$I$_{0.9}$)$_3$ in a magnetic field $B$ parallel to the
$c$ direction. Data are shifted consecutively by 0.2~J/molK$^2$ with
respect to $B = 0$.}
\label{ctvst10}
\end{figure}

\section{Experimental}
Single-crystalline samples of CsMn(Br$_x$I$_{1-x}$)$_3$ were grown by
the Bridgman technique at the Tokyo Institute of Technology \cite{ono98}.
For the measurements pieces of 24 to 111~mg were cleaved from the
crystals. The specific heat, $C$, was measured by a standard
semiadiabatic heat-pulse technique. Magnetic fields up to 14~T
were applied either along or perpendicularly to the clearly visible
$c$ axis of the crystals. The magnetocaloric effect, $(\delta T/
\delta B)_S = -(T/C)(\delta S/\delta B)_T$, was measured
in the same calorimeter. $S$ denotes the entropy of the system.
Upon changing the magnetic field by small steps $\Delta B$, the
resulting temperature variation $\Delta T$ was recorded. Taking
into account the small eddy-current heating the magnetocaloric
effect $\Delta T/\Delta B$ was extracted. For more details on the
experiment see Ref.\ \onlinecite{per98}.

\begin{figure}[ht]
  \centerline{\psfig{file=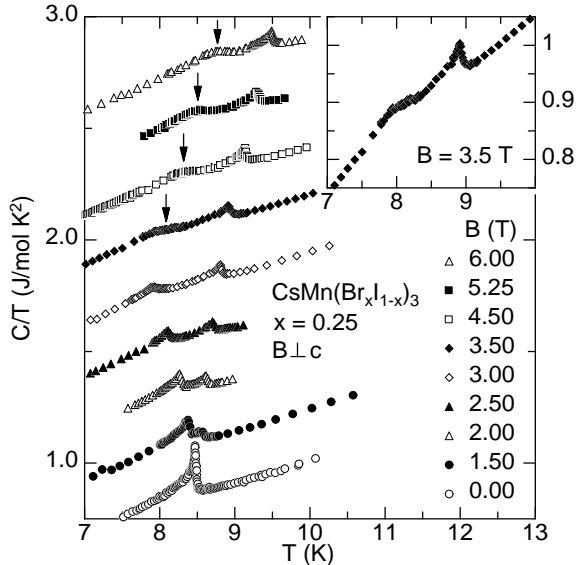,clip=,width=8cm}}
\caption[]{$C/T$ vs $T$ for CsMn(Br$_{0.25}$I$_{0.75}$)$_3$ in $B
\perp c$. Data are shifted consecutively by 0.2~J/molK$^2$ with
respect to $B = 0$. The arrows indicate the phase transitions between
the spin-flop phase and the intermediate collinear phase. The inset
shows an enlargement of the data at $B = 3.5$~T.}
\label{ctvst25}
\end{figure}

\section{Results and Discussion}
The specific heat of the sample with the smallest Br concentration,
CsMn(Br$_{0.1}$I$_{0.9}$)$_3$, is shown in Fig.\ \ref{ctvst10} for
different fields $B$ aligned along the $c$ direction. This easy-axis
system shows two consecutive zero-field transitions, which merge
into one at the spin-flop field of about 4~T. The steep anomaly
found beyond this field resembles that found for pure CsMnI$_3$
which was analyzed in terms of a chiral Heisenberg model at the
spin-flop field $B_M \approx 6.4$~T \cite{web95}. The substitution
of 10\% of the I$^-$ ions by Br$^-$ with the concomitant increase of
the (negative) anisotropy $D$ towards zero reduces considerably
the spin-flop field and likewise the width of the intermediate
phase ($T_{N2} = 8.36$~K $< T < T_{N1}$ = 9.80~K). This trend
continues further for a sample with $x = 0.18$ (data not shown)
where $T_{N1} = 8.50$~K, $T_{N2} = 8.40$~K, and $B_M \approx 1$~T
(see also Figs.\ \ref{phasdiag} and \ref{tn} below).

The data for $x = 0.25$, on the other hand, resemble those of the
pure easy-plane system CsMnBr$_3$ \cite{web95,gau89}, where a
magnetic field in the $ab$ plane quickly removes the chiral
degeneracy and leads to a splitting of the zero-field transition
(Fig.\ \ref{ctvst25}). In contrast to pure CsMnBr$_3$, however,
the anomaly at lower temperatures changes its appearance above
about 3~T, i.e., the anomaly (visualized by the arrows in
Fig.\ \ref{ctvst25}) becomes much more rounded and the feature
in $C$ shifts towards higher temperatures for increasing $B$
rather than to lower $T$ as in CsMnBr$_3$ \cite{web95,gau89}.
Indeed, what is reflected by the low-temperature anomalies in
Fig.\ \ref{ctvst25} are two different phase transitions;
from the chiral phase to the collinear phase at low $B$ and from
the spin-flop phase to the collinear phase at $B$ larger than about
3~T (see Fig.\ \ref{pds}(b) and also the phase diagram in
Fig.\ \ref{phasdiag} below). This result, therefore, reflects
the fact that the anisotropy $D$ for $x = 0.25$ has switched
from negative to positive, with $D < 3|J_{ab}|$. We found
a similar behavior with a considerably reduced width of the
collinear phase in the specific heat of a sample with $x = 0.20$
(data not shown).

\begin{figure}[ht]
  \centerline{\psfig{file=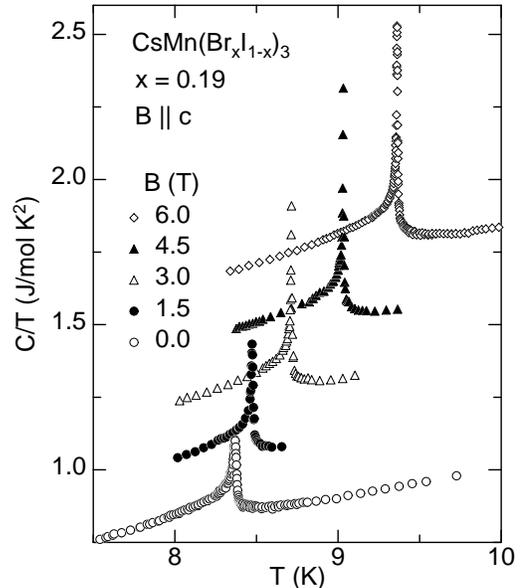,clip=,width=7.2cm}}
\caption[]{$C/T$ vs $T$ for CsMn(Br$_{0.19}$I$_{0.81}$)$_3$ in a magnetic
field $B$ parallel to the $c$ direction. Data are shifted consecutively by
0.2~J/molK$^2$ with respect to $B = 0$.}
\label{ctvst19p}
\end{figure}

Consequently, the anisotropy $D$ should become zero somewhere between
$x = 0.18$ and $x = 0.20$. A good candidate for such a chiral
Heisenberg system is therefore CsMn(Br$_{0.19}$I$_{0.81}$)$_3$.
For an isotropic Heisenberg system only one phase-transition
line from the paramagnetic to the chiral phase is expected,
independent of the magnetic-field orientation. Indeed,
magnetization and susceptibility data could not observe any
spin-flop line or splitting of the zero-field transition \cite{ono98}.
Our specific-heat data for $B \parallel c$ (Fig.\ \ref{ctvst19p})
are in line with these observations. We particularly can resolve only
a single strong anomaly at $T_N$ which becomes somewhat larger with
increasing field up to 6~T, similar to what is observed for $x = 0.1$
above $B_M$.

\begin{figure}[ht]
  \centerline{\psfig{file=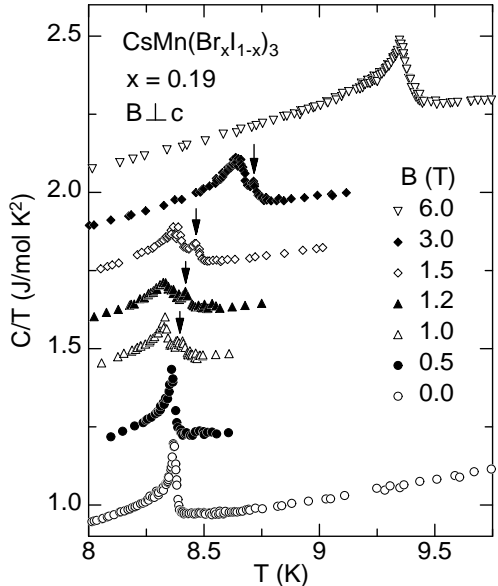,clip=,width=7cm}}
\caption[]{$C/T$ vs $T$ for CsMn(Br$_{0.19}$I$_{0.81}$)$_3$ in a magnetic
field $B$ perpendicular to the $c$ direction. Data are shifted
by different amounts with respect to $B = 0$. The arrows indicate
the small anomaly indicating the transition from the intermediate
phase to the paramagnetic phase.}
\label{ctvst19s}
\end{figure}

However, measurements of CsMn(Br$_{0.19}$I$_{0.81}$)$_3$ for $B \perp c$
(Fig.\ \ref{ctvst19s}) reflect a small residual planar anisotropy, as
evidenced by a slight splitting of the transition in fields between 1
and 3~T. The anomaly at lower temperatures is still relatively sharp
and large at $B = 1$~T, but becomes clearly reduced at 1.2~T which
indicates the junction with the spin-flop phase line. At higher
fields the phase lines merge and only one anomaly remains at 6~T.

\begin{figure}[ht]
  \centerline{\psfig{file=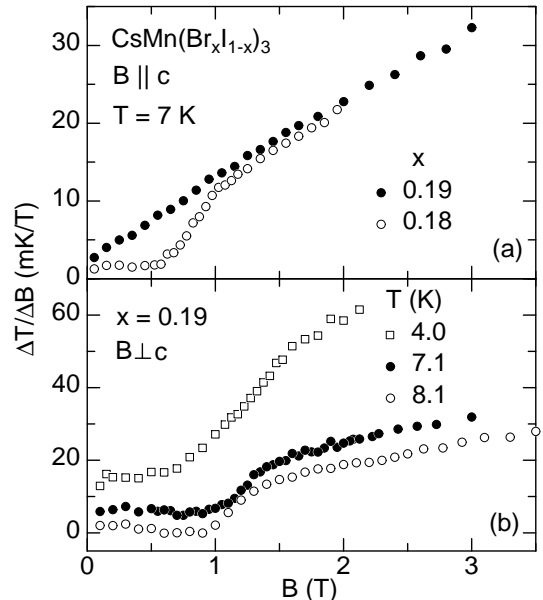,clip=,width=7.5cm}}
\caption[]{Magnetocaloric effect of (a) CsMn(Br$_{0.19}$I$_{0.81}$)$_3$
and CsMn(Br$_{0.18}$I$_{0.82}$)$_3$ in $B \parallel c$ and (b) of
CsMn(Br$_{0.19}$I$_{0.81}$)$_3$ for three different temperatures
in $B \perp c$. Data in (b) are shifted consecutively by 5~mK/T with
respect to $T = 8.1$~K.}
\label{mk19}
\end{figure}

In order to determine the complete phase diagrams including the
expected spin-flop lines (see Fig.\ \ref{pds}) we measured the
magnetocaloric effect for all samples. Since the spin-flop transition
at $B_c$ is almost temperature independent, the specific heat is not
sensitive to this transition, contrary to magnetocaloric-effect
measurements which cross the corresponding phase line
at an approximately right angle. Figure \ref{mk19}(a) shows
the magnetocaloric effect for the samples with $x = 0.18$ and $x =
0.19$ at $T \approx 7$~K in fields aligned parallel to the $c$ axis.
For $x = 0.18$, a clear step at about 0.9~T is visible which
signals the spin-flop transition in line with the data of Ono
et al.\ \cite{ono98}. The spin-flop field at each temperature
was estimated from the position of the maximum in the derivative
of the magnetocaloric-effect data. For $x = 0.19$, $\Delta T/\Delta B$
increases monotonically without any detectable step or anomaly.
This confirms that CsMn(Br$_{0.19}$I$_{0.81}$)$_3$ has
no Ising-like anisotropy.

Instead the anisotropy $D$ has switched to an $XY$ type, as
the specific-heat data (Fig.\ \ref{ctvst19s}) show. Consequently,
a spin-flop line is expected for fields perpendicular to $c$
[see Fig.\ \ref{pds}(b)]. Indeed, magnetocaloric-effect data
could verify this phase diagram by showing a step-like feature
at about 1.2~T almost independent of temperature [Fig.\ \ref{mk19}(b)].
Therefore, the critical concentration for which $D = 0$ should
be just below $x = 0.19$.

\begin{figure}[ht]
  \centerline{\psfig{file=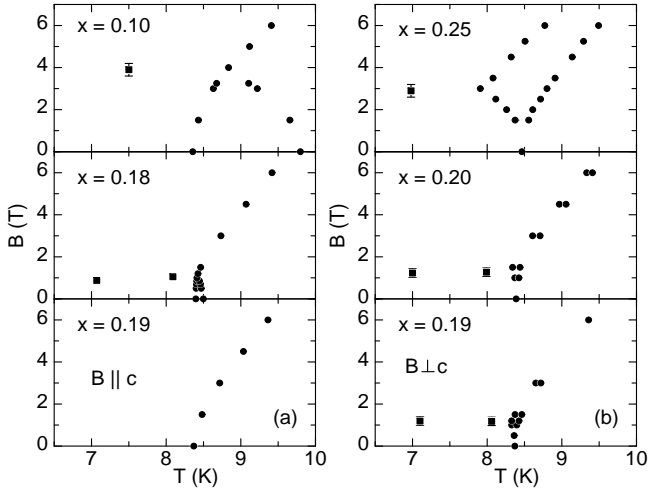,clip=,width=8.6cm}}
\caption[]{Phase diagrams of CsMn(Br$_x$I$_{1-x}$)$_3$ (a) for
fields along the $c$ direction for $x \leq 0.19$ and (b) for fields
perpendicular to $c$ for $x \geq 19$.}
\label{phasdiag}
\end{figure}

Figure \ref{phasdiag} summarizes the results in terms of $(B,T)$ phase
diagrams for various $x$. In Fig.\ \ref{phasdiag}(a), the absolute
magnitude of the easy-axis anisotropy $(D < 0)$ decreases with
increasing $x$, getting close to zero for $x = 0.18$. From the
reduced spin-flop field $B_c \approx 1$~T for $x= 0.18$ one can
estimate with Eq.\ (\ref{spinflop}) that $|J_cD|$ has reduced to
about 2.5\% of the value for CsMnI$_3$. Since $J_c$ should depend
little on $x$, this means that $|D|$ has reduced to about 95~neV
corresponding to 1.1~mK. The phase diagrams for
$x = 0.10$ and $x = 0.18$ fully agree with the predicted behavior
for easy-axis systems [Fig.\ \ref{pds}(a)] \cite{col97}.

The phase-diagram topology changes for easy-plane systems with $D > 0$.
In Fig.\ \ref{phasdiag}(b), the absolute magnitude of $D$ increases
with $x$, with a small anisotropy present for $x = 0.19$. As for $B
\parallel c$, the phase diagrams for $B \perp c$ $(D > 0)$ are in
full agreement with mean-field calculations and verify nicely the
predictions for easy-plane systems with $D < 3|J_{ab}|$ \cite{plu89}.

\begin{figure}[ht]
  \centerline{\psfig{file=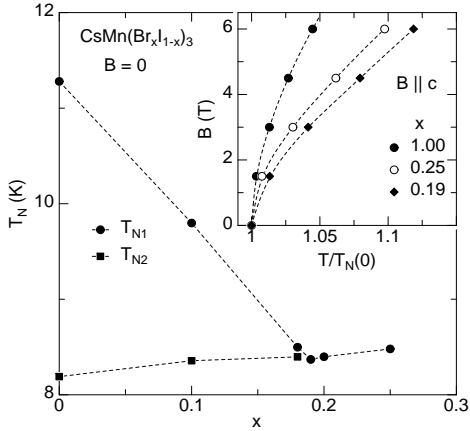,clip=,width=6.5cm}}
\caption[]{Transition temperatures $T_{N1}$ and $T_{N2}$ of
CsMn(Br$_x$I$_{1-x}$)$_3$ at $B = 0$ as a function of $x$. The data
for CsMnI$_3$ are from \cite{bec93a}. Lines are guide to the eye.
The inset shows the field dependence of the normalized transition
temperatures for three concentrations in $B \parallel c$.}
\label{tn}
\end{figure}

The phase diagram of the transition temperatures vs Br concentration
$x$ is shown in Fig.\ \ref{tn}. The lower N\'eel temperature $T_{N2}$
increases slightly with $x$, whereas $T_{N1}$ rapidly decreases. The two
phase lines merge at a critical concentration, $x_c$, somewhere between
$x = 0.18$ and $x = 0.19$. The phase diagram is in line with that
reported by Ono et al.\ \cite{ono98}. With our specific-heat and
magnetocaloric-effect measurements, however, we were able to resolve
the spin-flop line and the intermediate collinear phase for $x = 0.19$
at fields between 1 and 3~T proving that the critical concentration
with $D = 0$ must be slightly less than $x = 0.19$.

For completeness, the inset of Fig.\ \ref{tn} shows a comparative
$B-T$ phase diagram for different chiral $XY$ systems with $B$ aligned
along the $c$ direction. The data for CsMnBr$_3$ are from \cite{deu91}.
With increasing field the phase transition from the paramagnetic
to the chiral phase shifts to higher temperatures. This effect
is less prominent for larger $x$ indicating that the
field-induced $T_N$ increase becomes larger for reduced $D$.

\begin{figure}[ht]
  \centerline{\psfig{file=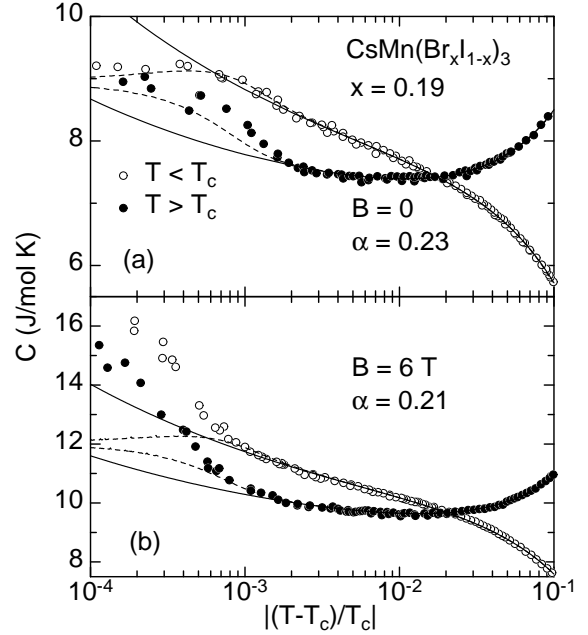,clip=,width=8cm}}
\caption[]{Specific heat $C$ of CsMn(Br$_{0.19}$I$_{0.81}$)$_3$
vs $\log|(T-T_c)/T_c|$ for (a) $B = 0$ and (b) $B = 6$~T. The solid
lines are fits according to Eq.\ (\ref{critexp}), the dash-dotted lines
are fits including a Gauss-distributed smearing of $T_c$.}
\label{kriexp}
\end{figure}

As a final point, we discuss the critical behavior of
CsMn(Br$_x$I$_{1-x}$)$_3$. In order to describe the specific-heat
data close to the critical temperature $T_c$ we applied the usual
fit function \cite{kor73}
\begin{equation}
C^{\pm} = (A^\pm/\alpha)|t|^{-\alpha} + B + Et,
\label{critexp}
\end{equation}
where $t = (T - T_c)/T_c$ and the superscript $+$ ($-$) refers to
$t > 0$ ($t < 0$). The first term describes the leading contribution
to the singularity in $C$ and the non-singular contribution to the
specific heat is approximated by $B + Et$. After a good fit of the data
had been achieved except very close to $T_c$, a Gaussian distribution
of $T_c$ with width $\delta T_c$ was introduced. This procedure is
able to describe a rounding of the transitions caused by sample
inhomogeneities (see also Refs.\ \cite{web95,loe95}).

Figure \ref{kriexp}(a) shows the specific heat $C$ for $x = 0.19$ vs
the reduced temperature $|t|$ at $B = 0$. The data are compatible with
an exponent $\alpha = 0.23(7)$ of the specific heat if we include a
Gaussian broadening of $\delta T_c/T_c \approx 4.2 \times 10^{-4}$.
The dashed lines indicate the fit with these parameters, while the
solid lines represent a fit with $\delta T_c = 0$. The exponent
$\alpha$ as well as the experimental amplitude ratio $A^+/A^- = 0.54(13)$
are in line with the chiral Heisenberg model, which predicts
$\alpha = 0.24(8)$ and $A^+/A^- = 0.54(20)$ \cite{kaw98}. This is
in accordance with the small easy-plane anisotropy which obviously
is too small to force the system to $XY$ chirality. The available
results of neutron-scattering experiments for $x = 0.19$ are rather
inconclusive \cite{ono98b}. While the exponent $\beta = 0.28(2)$ of
the sublattice magnetization agrees well with the theoretical value
of a chiral Heisenberg system ($\beta = 0.30$), the exponents of the
susceptibility, $\gamma = 0.75(4)$, and of the correlation length,
$\nu = 0.42(3)$, are at variance with the predictions ($\gamma =
1.17$ and $\nu = 0.59$). Furthermore, the experimentally found
exponents are in contradiction to the fundamental scaling laws
$\alpha + 2\beta + \gamma = 2$ and $\alpha + d\nu = 2$ ($d$ is the
dimension), which should be fulfilled at universal second-order phase
transitions.

Theoretically, the region around the multicritical point where a
chiral Heisenberg scenario is expected should not be large
\cite{kaw93}. This would imply that either the increase of the Br
concentration ($x > x_c$) or the application of a large magnetic
field ($B > B_M$) should drive the system quickly to chiral
$XY$ behavior with $\alpha = 0.34(6)$, which for CsMnBr$_3$ ($x = 1$)
has been observed \cite{wan91,deu92}. Nevertheless, for $x = 0.20$
and for $x = 0.25$ the critical exponents remain approximately
constant with $\alpha = 0.25(7)$ and $\alpha = 0.20(6)$, respectively,
suggesting that the chiral Heisenberg behavior is rather stable.

Another unexpected result becomes obvious from the analysis of the
specific-heat data of $x = 0.19$ in $B = 6$~T applied along the
$c$ direction [Fig.\ \ref{kriexp}(b)]. A magnetic field of this
strength, i.e., much larger than $B_M$, should induce $XY$ chirality
for a system with Ising anisotropy as previously observed for
CsNiCl$_3$ \cite{bec93b,end94} and, in this work, with $\alpha =
0.37(10)$ for CsMn(Br$_{0.1}$I$_{0.9}$)$_3$ at $B = 6$~T (not shown).
Likewise, for an easy-plane system like CsMnBr$_3$ the critical
behavior remains chiral $XY$-like for fields applied along $c$
\cite{deu92}. However, for $x = 0.19$ (as well as for $x = 0.18$,
data not shown) in a magnetic field of 6~T, $C$ for $t \ge 10^{-3}$
still follows a chiral Heisenberg-like behavior with $\alpha =
0.21(8)$ [$\alpha = 0.23(6)$ for $x = 0.18$] and increases much
more strongly for $t \rightarrow 0$ which can neither be described
by a reasonable critical exponent nor by a $T_c$ distribution. This
possibly indicates a crossover to a weakly first-order transition,
similar as previously observed for CsCuCl$_3$ close to $T_c$ at
$B = 0$ \cite{web96}.

In conclusion, we have mapped out in detail the impact of an
axial versus a planar anisotropy on the magnetic phase diagrams of
triangular-lattice antiferromagnets by fine-tuning $x$ of the
system CsMn(Br$_x$I$_{1-x}$)$_3$. In particular, the predicted
phase diagram [Fig.\ \ref{pds}(b)] of easy-plane systems with
small anisotropy could by accurately verified. The critical
concentration for which the spin anisotropy vanishes was
found to be located between $x = 0.18$, a system with small
axial anisotropy ($D < 0$), and $x = 0.19$, a system with small
planar anisotropy ($D > 0$). The critical behavior at $B = 0$
for $x = 0.19$, $x = 0.20$, and $x = 0.25$ can be described with
critical exponents $\alpha$ as predicted from Monte-Carlo
simulations for the chiral Heisenberg universality class \cite{kaw98}.
Thereby, rounding effects due to sample inhomogeneities prevent
the possible detection of a crossover to a first-order scenario
as proposed recently \cite{tis00}. In a magnetic field $B = 6$~T,
the samples with $x = 0.18$ and $x = 0.19$ show a weakly
first-order phase transition. For $10^{-3} < t < 0.1$, the data
can be described by chiral Heisenberg critical exponents. For all
other samples with either larger planar or larger axial symmetry,
no indication for a first-order phase transition was detected.



\begin{references}

\bibitem{kaw85}
H. Kawamura, J. Phys. Soc. Jpn. {\bf 54}, 3220 (1985); {\it ibid.} 
{\bf 55}, 2095 (1986).

\bibitem{kaw98}
H. Kawamura, J. Phys. Condens. Matter {\bf 10}, 4707 (1998) and
references therein.

\bibitem{loi98}
D. Loison and K.\ D. Schotte, Eur. Phys. J. B {\bf 50}, 735 (1998).

\bibitem{tis00}
M. Tissier, B. Delamotte, and D. Mouhanna, Phys. Rev. Lett. {\bf 84},
5208 (2000).

\bibitem{loi00}
D. Loison and K.\ D. Schotte, Eur. Phys. J. B {\bf 14}, 125 (2000).

\bibitem{web96}
H.\ B. Weber, T. Werner, J. Wosnitza, H. v. L\"ohneysen, and U. Schotte,
Phys. Rev. B {\bf 54}, 15\,924 (1996).

\bibitem{col97}
M.\ F. Collins and O.\ A. Petrenko, Can. J. Phys. {\bf 75}, 605 (1997).

\bibitem{kad88}
H. Kadowaki, S.\ M. Shapiro, T. Inami, and Y. Ajiro, J. Phys. Soc.
Jpn. {\bf 57}, 2640 (1988).

\bibitem{mas89}
T.\ E. Mason, B.\ D. Gaulin, and M.\ F. Collins, Phys. Rev. B {\bf 39},
586 (1989).

\bibitem{wan91}
J. Wang, D.\ P. Belanger, and B.\ D. Gaulin, Phys. Rev. B {\bf 66},
3195 (1991).

\bibitem{deu92}
R. Deutschmann, H. v. L\"ohneysen, J. Wosnitza, R.\ K. Kremer, and
D. Visser, Europhys. Lett. {\bf 17}, 637 (1992).

\bibitem{pla00}
V.\ P. Plakhty, J. Kulda, D. Visser, E.\ V. Moskvin, and J. Wosnitza,
Phys. Rev. Lett. {\bf 85}, 3942 (2000).

\bibitem{web95}
H. Weber, D. Beckmann, J. Wosnitza, H. v. L\"ohneysen, and D. Visser,
Int. J. Mod. Phys. B {\bf 12}, 1387 (1995).

\bibitem{loe95}
H. v. L\"ohneysen, D. Beckmann, J. Wosnitza, and D. Visser,
J. Magn. Magn. Mat. {\bf 140-144}, 1469 (1995).

\bibitem{bec93b}
D. Beckmann, J. Wosnitza, H. v. L\"ohneysen, and D. Visser,
Phys. Rev. Lett. {\bf 71}, 2829 (1993).

\bibitem{end94}
M. Enderle, G. Furtuna, and M. Steiner, J. Phys. Condens. Matter
{\bf 6}, L385 (1994).

\bibitem{wos94}
J. Wosnitza, R. Deutschmann, H. v. L\"ohneysen, and R.\ K. Kremer,
J. Phys. Condens. Matter {\bf 6}, 8045 (1994).

\bibitem{ono98}
T. Ono, H. Tanaka, T. Kato, and K. Iio, J. Phys. Condens. Matter {\bf 10},
7209 (1998).

\bibitem{har91}
A. Harrison, M.\ F. Collins, J. Abu-Dayyah, C.\ V. Stager, Phys. Rev. B
{\bf 43}, 679 (1991).

\bibitem{bec93a}
D. Beckmann, J. Wosnitza, H. v. L\"ohneysen, and D. Visser,
J. Phys. Condens. Matter {\bf 5}, 6289 (1993).

\bibitem{plu88}
M.\ L. Plumer, K. Hood, and A. Caill\'e, Phys. Rev. Lett. {\bf 60},
45 (1988).

\bibitem{kaw90}
H. Kawamura, A. Caill\'e, and M.\ L. Plumer, Phys. Rev. B {\bf 41},
4416 (1990).

\bibitem{plu89}
M.\ L. Plumer, A. Caill\'e, and K. Hood, Phys. Rev. B {\bf 39}, 4489
(1989).

\bibitem{gau89}
B.\ D. Gaulin, T.\ E. Mason, M.\ F. Collins, and J.\ Z. Larese,
Phys. Rev. Lett. {\bf 62}, 1380 (1989).

\bibitem{per98}
F. P\'erez, T. Werner, J. Wosnitza, H. v. L\"ohneysen, and H. Tanaka,
Phys. Rev. B {\bf 58}, 9316 (1998).

\bibitem{deu91}
R. Deutschmann, J. Wosnitza, and H. v. L\"ohneysen, unpublished.

\bibitem{kor73}
A. Kornblit and G. Ahlers, Phys. Rev. B {\bf 8}, 5163 (1973).

\bibitem{ono98b}
T. Ono, H. Tanaka, T. Kato, K. Iio, K. Nakajima, and K. Kakurai,
J. Magn. Magn. Mat. {\bf 177-181}, 735 (1998);
T. Ono, H. Tanaka, T. Kato, K. Nakajima, and K. Kakurai,
J. Phys. Condens. Matter {\bf 11}, 4427 (1999).

\bibitem{kaw93}
H. Kawamura, Phys. Rev. B {\bf 47}, 3415 (1993).

\end{references}
\end{document}